\documentclass[showpacs,reprint,twocolumn,floatfix]{revtex4-1}

\usepackage{graphicx,float,amsmath}
\usepackage{natbib}
\usepackage{lipsum}

\begin{document}

\title{Giant, anomalous piezo-impedance of silicon-on-insulator}

\author{H. Li$^1$}
\author{C.T.K. Lew$^2$}
\author{B.C. Johnson$^2$}
\author{J.C. McCallum$^3$}
\author{S. Arscott$^4$}
\author{A.C.H. Rowe$^1$}
\email{alistair.rowe@polytechnique.edu}

\affiliation{$^1$Physique de la Mati\`ere Condens\'ee, Ecole Polytechnique, CNRS, Universit\'e Paris-Saclay, 91128 Palaiseau, France}
\affiliation{$^2$Centre for Quantum Computation \& Communication Technology, School of Physics, University of Melbourne, VIC 3010, Australia}
\affiliation{$^3$School of Physics, University of Melbourne, Melbourne, Victoria 3010, Australia  }
\affiliation{$^4$Institut d'Electronique, de Micro\'electronique et de Nanotechnologie (IEMN), Universit\'e de Lille, CNRS, Avenue Poincar\'e, Cit\'e Scientifique, 59652 Villeneuve d'Ascq, France}

\begin{abstract}
A giant, anomalous piezo-response of fully-depleted silicon-on-insulator (FD-SOI) devices under mechanical stress is demonstrated using impedance spectroscopy. This piezo-response strongly depends on the measurement frequency, $\omega$, and consists of both a piezoresistance (PZR) and piezocapacitance whose maximum values are $\pi_R = -1100 \times 10^{-11}$ Pa$^{-1}$ and $\pi_C = -900 \times 10^{-11}$ Pa$^{-1}$ respectively. These 
values should be compared with the usual bulk PZR in p-type silicon, $\pi_R= 70 \times 10^{-11}$ Pa$^{-1}$. The observations are well described using models of space charge limited electron and hole currents in the presence of fast electronic traps having stress-dependent capture ($\omega_c$) and emission rates. Under steady-state conditions (i.e. when $\omega \ll \omega_c$) where the impedance spectroscopy measurements yield results that are directly comparable with previously published reports of PZR in depleted, silicon nano-objects, the overall piezo-response is just the usual, bulk silicon PZR. Anomalous PZR is observed only under non-steady-state conditions when $\omega \approx \omega_c$, with a symmetry suggesting that the electro-mechanically active fast traps are native Pb$_0$ interface defects. The observations suggest new functionalities for FD-SOI, and shed light on the debate over the PZR of carrier depleted nano-silicon.

\end{abstract}
\maketitle

\section{Introduction}
\label{Intro}

Mechanical stress modifies the electronic structure of solids and gives rise to a change in electrical resistivity, $\Delta\rho$, known as the piezoresistance (PZR) \cite{smith1954}. The PZR, which in crystalline solids may be a tensor quantity, is characterized by a $\pi$-coefficient: \begin{equation} \label{pi} \pi = \frac{1}{X}\frac{\Delta\rho}{\rho_0}, \end{equation} where $X$ is the applied stress and $\rho_0$ is the zero-stress resistivity. In solids whose electronic structure is well described by a simple, spherical band model, the PZR is due mainly to stress-induced changes in the atomic density, and hence to the equilibrium density of free charge carriers \cite{sagar1958,rowe2014}. However, if the electronic structure consists of multiple degenerate bands at a single point in the Brillouin zone, or multiple valleys, then the stress-induced density changes are typically negligible compared to changes in the density-of-states weighted effective masses. The PZR is then principally determined by stress-induced changes to the charge carrier mobility. This is the case for silicon where stress-induced inter-valley charge transfer results in large PZR for n-type material \cite{smith1954,herring1956}, and interchange of heavy- and light- holes yields similarly large PZR in p-type material \cite{smith1954,milne2012}. The latter case is widely exploited in strained-silicon finFETs \cite{hoyt2002} in order to symmetrize the transconductance gain of the n-channel and p-channel transistors in CMOS circuitry \cite{lee2006}. One technologically important case of relevance here is that of p-type silicon in which both current and stress are parallel to the $\langle 110 \rangle$ crystal direction, in which case $\pi_{\textrm{bulk}} \approx 70\times10^{-11}$ Pa$^{-1}$.

Unlike bulk crystals, silicon nano-objects such as nanowires \cite{simpkins2008}, can be strongly depleted of free charge carriers by reducing the doping level to a point where the surface depletion layer width, $W$, is larger than a characteristic dimension of the object, $t$. In this space charge limited current (SCLC) regime, unusual and varied PZR has been reported, including giant effects up to $\pi \approx 3500 \times 10^{-11}$ Pa$^{-1}$ in nanowires \cite{he2006, neuzil2010} and nanomembranes \cite{yang2010, yang2011}, and anomalous (i.e. negative) effects comparable in magnitude to $\pi_{\textrm{bulk}}$ in nanowires \cite{lugstein2010, winkler2015} and nanomembranes \cite{jang2014}. Many others however, report PZR similar in both sign and magnitude to $\pi_{\textrm{bulk}}$ \cite{toriyama2003, reck2008, bui2009, milne2010, mile2010, barwicz2010, koumela2011, kumar2013, vietadao2015, mcclarty2016}, even when $W \gg t$. While some studies are made at high stresses where non-linearities may be important \cite{lugstein2010, kumar2013, winkler2015}, the majority are made with $X <$ 100 MPa, so it is unclear why such a variety of different behaviors have been observed in nominally very similar nano-objects \cite{rowe2014}. 

Importantly, all prior experiments \cite{smith1954, sagar1958, he2006, neuzil2010, yang2010, yang2011,lugstein2010,winkler2015,jang2014,toriyama2003, reck2008, bui2009, milne2010, mile2010, barwicz2010, koumela2011, kumar2013, vietadao2015, mcclarty2016} used DC methods to measure the PZR. In this approach a DC voltage, $V_{ds}$, is applied between the source and drain contacts of a device, and its resistance is estimated by measuring the resulting current, $I_0$. The stress-induced change in the current, $\Delta I$, for a fixed applied voltage, can then be used to estimate the $\pi$-coefficient in Eq. (\ref{pi}) according to $\pi \approx -1/X \times\Delta I/I_0$. While this approach has the advantage of simplicity, it misses important physical effects as will be shown here. In this work a different approach, based on impedance spectroscopy, is used. 

Impedance spectroscopy -- the study of electrical impedance as a function of frequency, $\omega$, is widely employed in solids, liquids, or at interfaces where a space charge is present, because it reveals details of charge relaxation dynamics and transport kinetics \cite{barsoukov2018}. It is particularly well established in electrochemistry \cite{orazem2011}, but has also found use in organic semiconductors where the SCLC regime is often encountered \cite{martens1999}, and to characterize the SCLC in photo-excited p-n junctions fabricated from both traditional inorganic semiconductors \cite{mihailetchi2005, mora2009} as well as more novel materials \cite{pockett2015}. 

The principal strength of impedance spectroscopy compared to the DC method for estimating PZR, is that it allows for a measurement of the $\pi$-coefficient under both \textit{steady-state} and \textit{non-steady-state} conditions, whereas the DC approach reveals only the steady-state PZR. Here it will be shown that in steady-state, impedance spectroscopy reveals that the SCLC in simple, resistor-like devices fabricated from fully-depleted silicon-on-insulator (FD-SOI)  exhibits a PZR comparable in sign and magnitude to the usual bulk effect \cite{smith1954} in agreement with Refs. \onlinecite{toriyama2003, reck2008, bui2009, milne2010, mile2010, barwicz2010, koumela2011, kumar2013, vietadao2015, mcclarty2016} while giant, anomalous PZR is observed only under non-steady-state conditions. Analysis of the impedance as a function of $\omega$ reveals that the cross-over between steady-state and non-steady state conditions is determined by the capture and emission rates of fast electronic traps associated with crystal defects. The giant, anomalous PZR is therefore ascribed to stress-induced changes in these capture and emission rates which give rise to stress-induced changes in the \textit{non-equilibrium} electron and hole densities. Moreover, the stress-dependent density of trapped charge gives rise to a giant piezocapacitance (PZC), another phenomenon which is inaccessible using DC methods, and is only revealed thanks to impedance spectroscopy.

\begin{figure}[t]
\includegraphics[clip,width=7.5 cm] {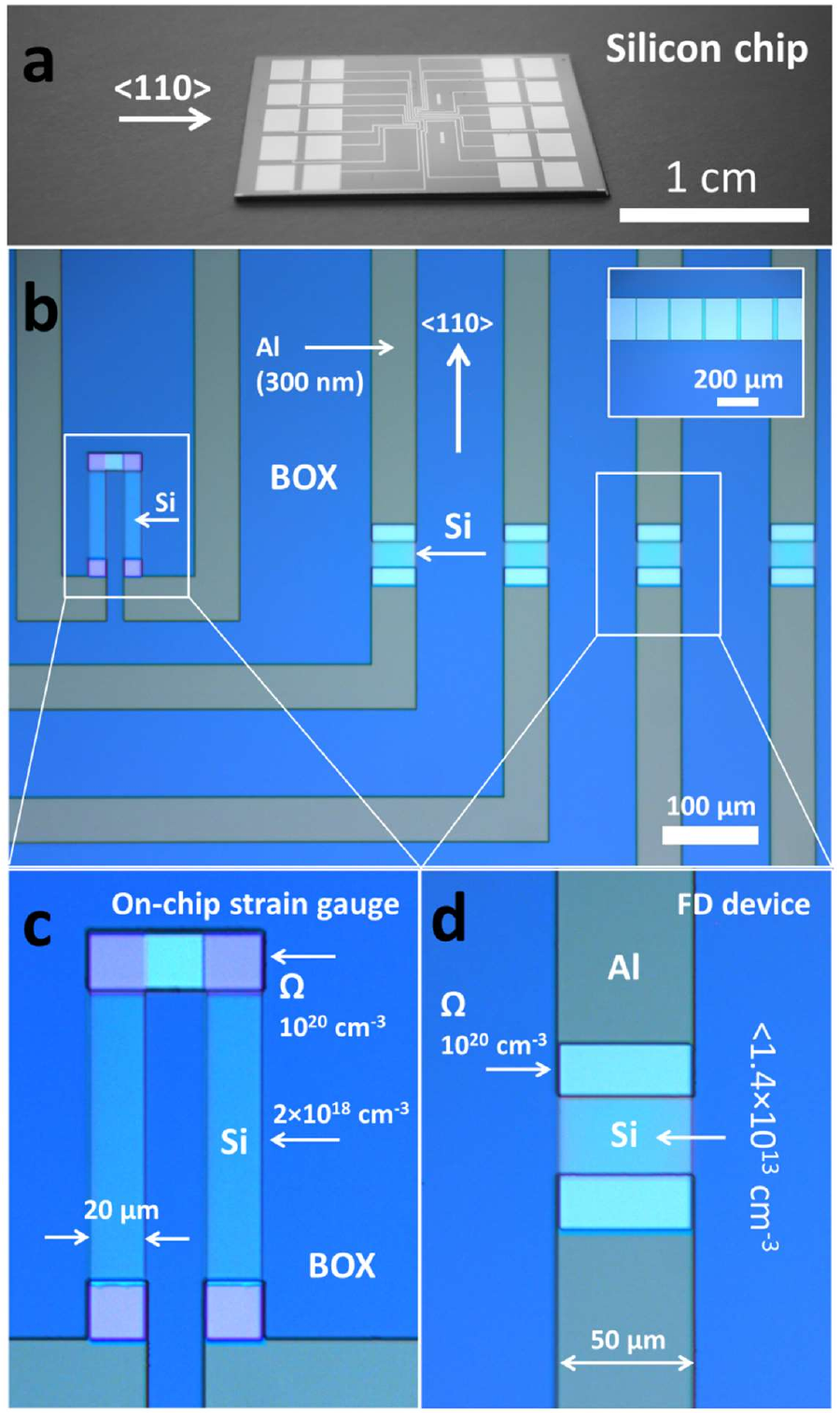}
\caption{(a) Photograph of an example silicon chip fabricated for the study. The large contact pads (2 mm $\times$ 2 mm) are visible on the chip which measures 1.3 cm by 2 cm. (b) A zoomed microscope image taken at the center of the chip shows the U-shaped strain gage and four FD devices. The inset shows a resistance ladder used for the measurement of the specific contact resistivity. (c) The on-chip strain gauges. (d) The fully-depleted silicon devices. Current flows in the devices and in the on-chip strain gages parallel to the $\langle 110 \rangle$ direction. This is also the direction along which uni-axial stress is applied.}
\label{sample}
\end{figure}

\section{Sample and measurement details}

Standard photo-lithographic processing methods are used to produce samples from the $t = 2$ $\mu$m thick device layer of a 3-inch, $(001)$-oriented FD-SOI wafer (buried oxide or BOX thickness, 2 $\mu$m, and a handle thickness of 400 $\mu$m). The device layer and handle are non-intentionally-doped with boron, $p < 1.4 \times 10^{13}$ cm$^{-3}$. After processing, the wafer is cut to form macroscopically large chips whose long axis is parallel to the $\langle 110 \rangle$ crystal direction as shown in Fig. \ref{sample}(a). As indicated in Fig. \ref{sample}(b), each chip contains four fully-depleted (FD) devices whose  active area is 50 $\mu$m wide and 30 $\mu$m long (see Fig. \ref{sample}(d)), and which have ohmic contacts and lines that run to large area pads (2 mm $\times$ 2 mm, visible in Fig. \ref{sample}(a)) used for external contacting. In addition to the FD devices, each chip also contains a silicon strain gage shown in Fig. \ref{sample}(c) and a resistance ladder, shown inset in Fig. \ref{sample}(b).

The resistance ladders are use to measure the specific contact resistivity, $\rho_s$, of the ohmic contacts made to the gages and FD devices using boron implantation ($p = 10^{20}$ cm$^{-3}$) followed by metallization and a 450 $^{\circ}$C post-anneal. This yields contacts with $\rho_s = 1.6 \times 10^{-6}$ $\Omega$ cm$^2$, sufficiently low that they do not contribute to the 2-terminal resistances of either the strain gages or the FD devices. As indicated in Fig. \ref{sample}(c), the gages are formed by uniformly implanting the active arms (vertical in the image) with a boron density of $2 \times 10^{18}$ cm$^{-3}$, and by fabricating a metallic short perpendicular to the active arms, yielding a fully ohmic device whose PZR is given by $\pi_{\textrm{bulk}}$. The gages can then be used to estimate \textit{in situ} the $\langle 110 \rangle$-oriented uni-axial applied stress near the center of the chip. 

\begin{figure}[t]
\includegraphics[clip,width=8.5 cm] {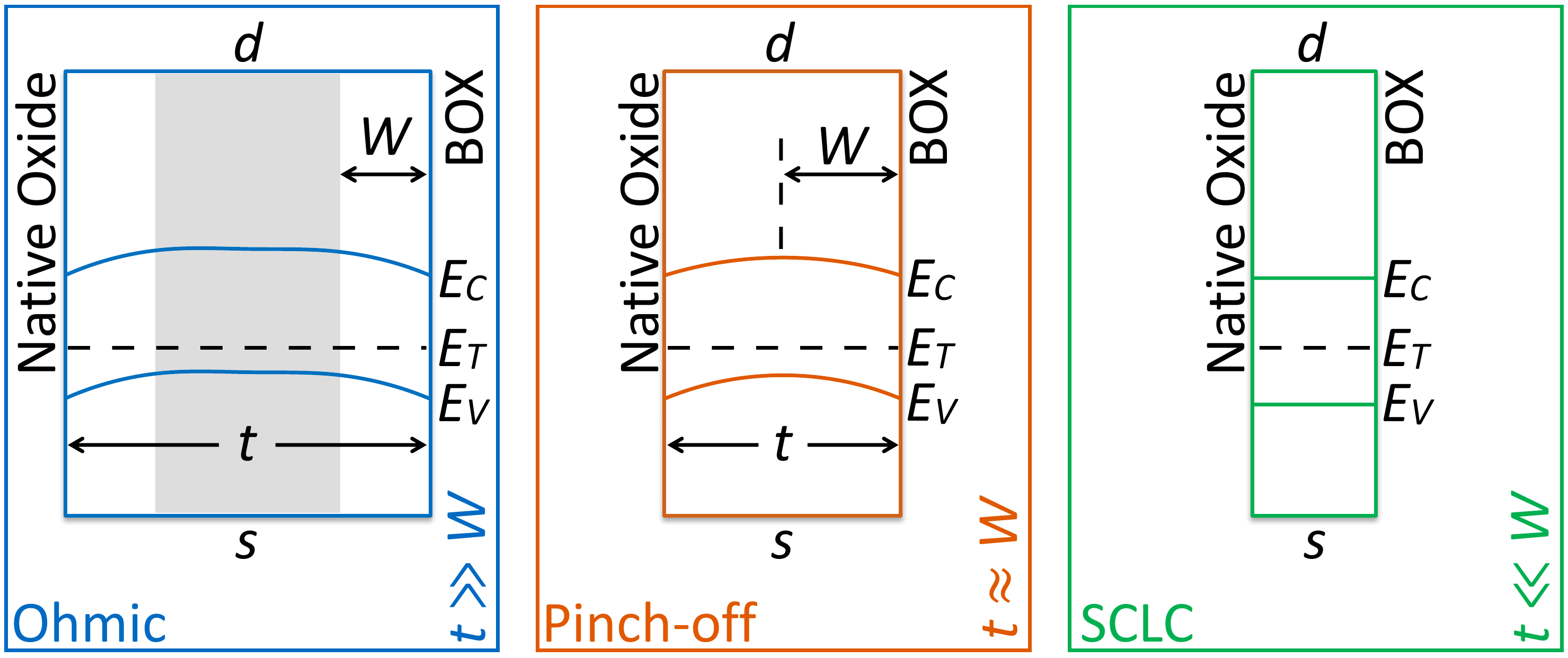}
\caption{A schematic side-view of the device active area demonstrating the concept of full surface depletion \cite{rowe2008}. The active arms of the strain gages are sufficiently doped so that $W \ll t$ and conduction between the drain ($d$) and the source ($s$) is ohmic (blue box). In the FD devices $W \gg t$ so that band bending is absent, and the Fermi level throughout the device is determine by surface pinning at an energy $E_T$ determined by the interface traps. In this limit a SCLC is expected (green box).}
\label{depletion}
\end{figure}

\begin{figure}[t]
\includegraphics[clip,width=8 cm] {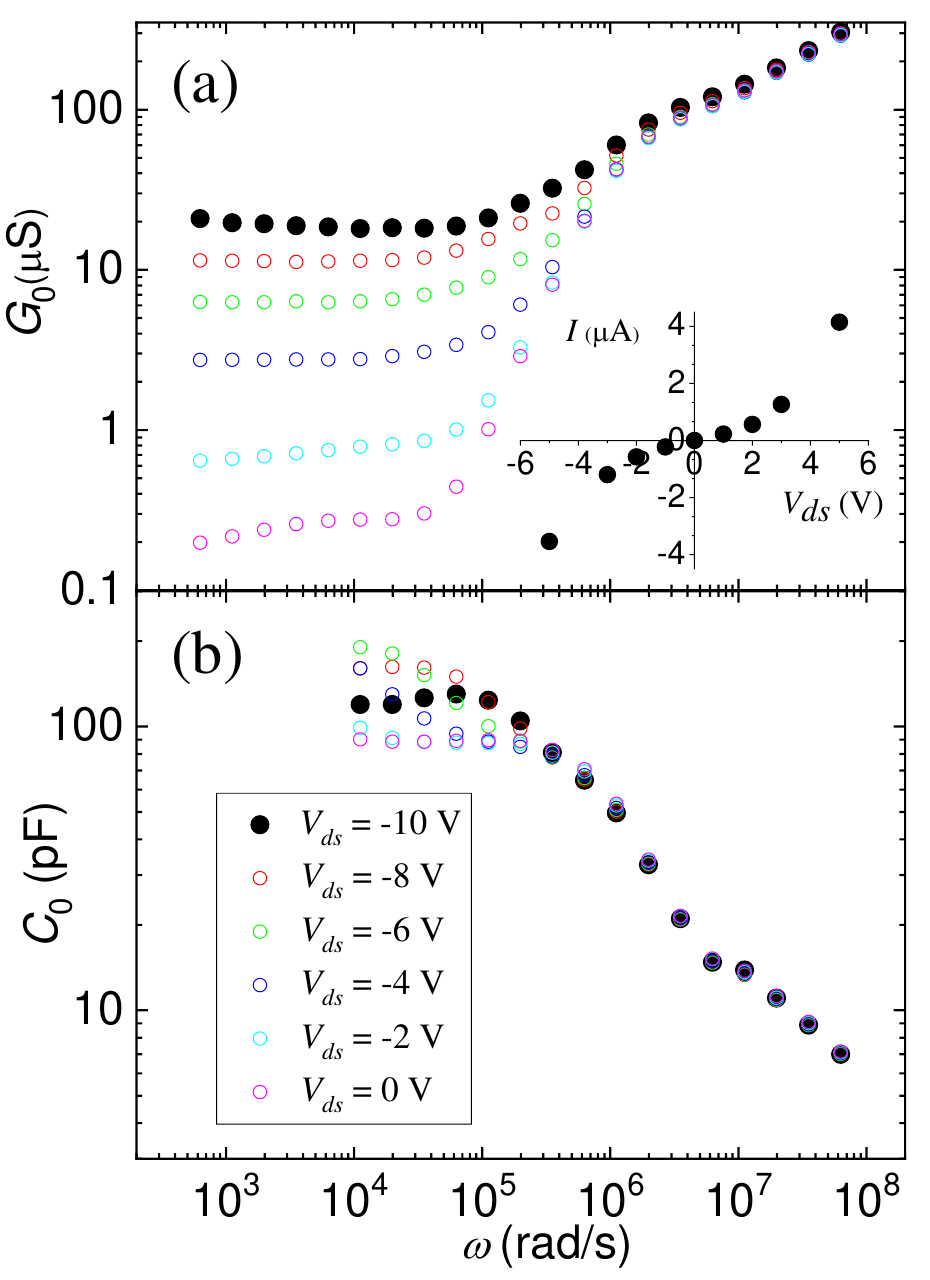}
\caption{Zero-stress measurements of $G_0$ (top) and $C_0$ (bottom) as a function of $\omega$ and $V_{ds}$. The inset (top) shows the steady-state current-voltage characteristic whose non-linearity is the first evidence of a SCLC. The frequency variation of $G_0$ and $C_0$ are consistent with a SCLC in the presence of fast traps\cite{kassing1975}. At low frequencies capacitance measurements are difficult and the data is not shown.}
\label{nostress}
\end{figure}

The FD devices themselves (shown in Fig. \ref{sample}(d)) are fabricated without modifying the background boron doping density of the FD-SOI device layer, and are the same type of device in which bulk, steady-state PZR was previously found using DC methods \cite{milne2010}. The low doping density of the device layer yields $W \gg t$ so that the ohmic, bulk channel which exists in the gages (in gray in Fig. \ref{depletion}) is pinched off, and the device layer is fully-depleted (see graphical argument in Fig. \ref{depletion}). The 2-terminal resistance is then dominated by the 30 $\times$ 50 $\mu$m active area and, like the strain gages, current flows parallel to the $\langle 110\rangle$ crystal direction. The steady-state current-voltage characteristics are non-linear despite the ohmic contacts (see inset, Fig. \ref{nostress}), a strong indication of the relative absence of free charge carriers in equilibrium. Charge transport then occurs in the SCLC regime due to double injection of non-equilibrium electrons and holes from the two ohmic contacts \cite{lampert1961}.

Mechanical stress is applied by clamping the left and right ends of the chip in Fig. \ref{sample}(a) and then pushing along the center-line of the chip i.e. using a three-point bending method. This method can be used to obtain both uni-axial tensile and compressive stresses of the order of several tens of MPa, whose magnitude is measured \textit{in situ} with the silicon strain gage as discussed above. In this way the piezo-response of the FD devices can be directly compared to that of the strain gages. It is also noted that the magnitude of the applied stress is modulated between zero and the desired, non-zero value at a frequency of the order of 0.2 Hz in order to avoid any measurement drift issues \cite{milne2010}.

In terms of the electrical measurement, in addition to the advantages presented in Section \ref{Intro}, impedance spectroscopy is also a natural choice for the estimation of the $\pi$-coefficient if the device characteristic is non-linear. This is because a $V_{ds}$-dependent resistance requires a measurement of the differential conductance. In the following, the in-phase and out-of-phase components of the current resulting from a total applied bias, $V(t) = V_{ds} + \exp(i\omega t)$, are assimilated with a conductance, $G$, and a capacitance, $C$, respectively. Both quantities may change with applied stress, $V_{ds}$ and $\omega$, so it is possible to define two $\pi$-coefficients, one for the PZR: \begin{equation} \label{piR} \pi_R\left(\omega,V_{ds}\right)\approx -\frac{1}{X}\frac{\Delta G}{G_0}, \end{equation}
where $\Delta G$ is the stress-induced change in $G$, and $G_0$ is the zero-stress conductance, and one for the PZC: \begin{equation} \label{piC} \pi_C\left(\omega,V_{ds}\right)\approx -\frac{1}{X}\frac{\Delta C}{C_0}, \end{equation} where $\Delta C$ is the stress-induced change in $C$, and $C_0$ is the zero-stress capacitance. These equations are valid for small relative changes in $G$ and $C$, and when stress-induced geometry changes are negligible. The overall piezo-impedance coefficient is then: 

\begin{widetext}
\begin{equation} \label{piZ} \pi_Z\left(\omega,V_{ds}\right)= \pi_R \frac{G_0^2}{G_0^2+\omega^2 C_0^2} + \pi_C \frac{\omega^2 C_0^2}{G_0^2+\omega^2 C_0^2} + i\left(\pi_C-\pi_R\right) \frac{\omega G_0 C_0}{G_0^2+\omega^2 C_0^2}. \end{equation}
\end{widetext}

The $\omega$-dependent measurements of $G$ and $C$ are made with a commercial impedance analyzer (HP 4192A LF) with the four-probe (16048A test leads) attachment. A conversion from the four-probe to the two-probe geometry is made approximately 10 cm from the sample holder according to rules outlined in the impedance analyzer user manual. This conversion is designed to minimize parasitic capacitances in the two-terminal part of the circuit. Importantly, the impedance analyzer's zero correction function is used to account for the conductance and capacitance of the external measurement circuit, thereby ensuring that the measured admittance is that of the device under test only. This is particularly important at frequencies in the 1 MHz to 10 MHz range where parasitic resonances can occur. Failure to perform these corrections can result in spurious estimates of the high frequency PZR and PZC.

\section{Zero stress FD device properties}
\label{zero}

Figure \ref{nostress} shows typical results for the $\omega$ and $V_{ds}$ dependence of $G_0$ and $C_0$. The increase in $G_0$ (decrease in $C_0$) at intermediate frequencies are signatures of small-signal SCLC in the presence of fast electronic traps associated with crystal defects \cite{kassing1975}. The values of $G_0$ and $C_0$ depend upon the device$’$s geometric capacitance ($C_g$) and conductance ($G_g$), the source-to-drain transit time ($T$), as well as the traps’ characteristic capture ($\omega_c$) and emission ($\omega_e$) rates. The strong voltage dependence of $G_0$ observed at low frequencies is due to the voltage dependence of $\omega_e$ and will be discussed further in Section \ref{volts}. The frequency dependence of both $G_0$ and $C_0$ may be estimated using Kassing’s model \cite{kassing1975} which provides an analytic solution to a simplified set of coupled differential equations (see Section \ref{KassingEquations}) that account for a single trap and carrier type. Although recombination is therefore absent in Kassing's model, it will be seen that despite its potentially limited applicability to the double injection case where multiple trap types are present (see discussion in Appendix \ref{KassingEquations}), the model is never-the-less useful in providing physical insight into the origin of the piezo-response data presented below. To demonstrate this as simply as possible, Kassing's model will be applied to the zero-stress and piezo-response data obtained for $V_{ds} =$ 10 V in the remainder of this work.  

\begin{figure*}[t]
\includegraphics[clip,width=16 cm] {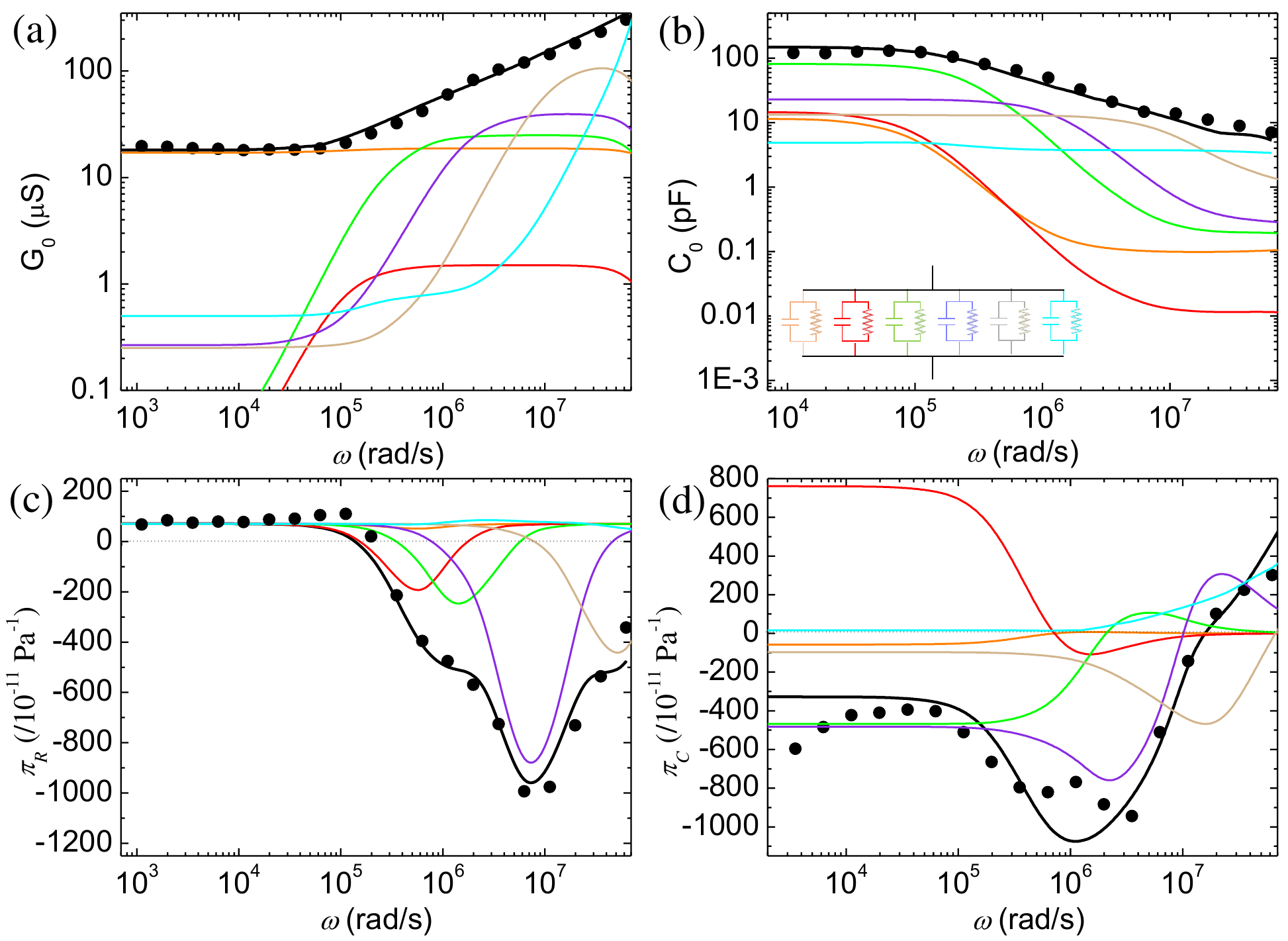}
\caption{Simultaneous modeling of zero-stress and piezo-response data (black dots) obtained as a function of $\omega$ for $V_{ds}$ = 10 V. Six individual trap types represented by the colored curves simulate a quasi-continuous trap distribution consistent with the PICTS signal in Appendix \ref{pictsappendix}. The individual results are added in parallel (see inset in (b)) to obtain the overall response (black curves). Fit parameters can be found in Tables \ref{zerostressparams} and \ref{piezoparams}.}
\label{Kassing}
\end{figure*}

Kassing’s model yields typical curves given by, for example, the purple lines in Figs. \ref{Kassing}(a) and \ref{Kassing}(b) for conductance and the capacitance respectively at $V_{ds}$ = 10 V. Notice that the calculated low-to-high frequency variation occurring in both quantities around $\omega \approx \omega_c$, is too rapid compared with the data. This is because the FD devices contain a continuous spectrum of traps as indicated by the asymmetric, broad hump centered at about 180 K in the photo-induced current transient spectroscopy (PICTS) signal shown in Fig. \ref{picts} of Appendix \ref{pictsappendix}. Broad PICTS signals are usually indicative of a continuum of traps whose spectrum is spread by disorder at the silicon/oxide interface \cite{papaioannou1989}.

A full description of the trap spectrum would therefore be cumbersome, so a compromise solution consisting of the inclusion of a finite number of different trap types is considered. Here six traps represented by the colored curves in Figs. \ref{Kassing}(a) and \ref{Kassing}(b) are introduced and, as an example of how to apply Kassing's model, the six individual values of $G$ and $C$ for each trap type are calculated for $V_{ds} = 10$ V using the trap parameters shown in Table \ref{zerostressparams}. A single transit time, $T = 2 \times 10^{-8}$ s, is used for all traps. Similarly, common values of the geometric conductance and capacitance are used for all traps, $G_g =10^{-4}$ S and $C_g = 6.5 \times 10^{-13}$ F. In each case the curve calculated from Kassing’s model is weighted by a pre-factor that represents the relative densities of each of the six traps. The overall values of $G_0$ and $C_0$ are then found by adding the individual conductance and capacitance values in parallel as indicated (inset) in Fig. \ref{Kassing}(b). It is interesting to note that in all cases except one (trap type number 1), the traps are fast ($\omega_e \ll \omega_c$), with the remaining case, depicted in orange in Fig. \ref{Kassing}, being slow i.e. $\omega_c < \omega_e$. As will be seen in Section \ref{piezo}, this is intimately related to the $V_{ds}$-dependence of the piezo-response. The choice of parameter values must simultaneously fit both $G_0$ and $C_0$ data, so the excellent agreement between the calculated values (black curves in Figs. \ref{Kassing}(a) and (b)) and the data (filled, black circles) is highly satisfactory.

\begin{table}[t]
    \caption{The fitting parameters used with Kassing's model \cite{kassing1975} to calculate $G_0$ in Fig. \ref{Kassing}(a), and $C_0$ in Fig. \ref{Kassing}(b).}
    \centering
    \begin{tabular}{c|c|c|c|c}
      \textbf{Trap} & \textbf{Line color} & $\omega_c$ & $\omega_e$ & \textbf{Relative}\\
       \textbf{number} & \textbf{in Fig. \ref{Kassing}} & (rad/s) & (rad/s) & \textbf{weight} \\
      \hline
      1 & orange & 16875 & 87750 & 0.204\\
      2 & red & $10^5$ & 100 & 0.03\\
      3 & green & $3 \times 10^5$ & 100 & 0.5\\
      4 & purple & $1.5 \times 10^6$ & 5000 & 0.8\\
      5 & brown & $10^6$ & $10^4$ & 2.5\\
      6 & cyan & $10^7$ & $10^5$ & 5\\
    \end{tabular}
  \label{zerostressparams}
\end{table}

The qualitative picture which emerges is as follows. For $\omega \ll \omega_c$ the applied voltage period is long compared to the lifetime of injected carriers in the band ($2\pi/\omega_c$) so that injected, non-equilibrium charge fully relaxes from the band within a voltage cycle i.e. the free charge population reaches steady-state. Results obtained in this frequency range are therefore directly comparable with those obtained using DC methods. For fast traps it can be shown \cite{kassing1975} that the density of the injected charge remaining in the band in steady-state is approximately a factor $\omega_e/\omega_c \ll 1$ smaller than the trap-free case, and hence $G_0$ is small and frequency-independent, \begin{equation} \label{steadystate} G_0\left(\omega \ll \omega_c\right) \approx \frac{\omega_e G_g}{\omega_c} \ll G_g, \end{equation} as seen in Fig. \ref{nostress}. Eq. (\ref{steadystate}) will be important when considering the steady-state PZR in Section \ref{piezo}. Conversely, a fraction $\omega_c/(\omega_e+\omega_c)$ of the injected charge is trapped during the voltage cycle. It approaches 1 for fast traps and is closer to zero for slow traps. Any trapped charge increases $C_0$ relative to $C_g$, to $C_g/(\omega_c T)$ \cite{kassing1975}. 

For $\omega \gg \omega_c$, the voltage period is short compared to $2\pi/\omega_c$ so that relaxation of injected charge from the band is negligible; in effect the device behaves as if no traps were present. To within a factor of the order of unity, $G_0 \rightarrow G_g$ and $C_0 \rightarrow C_g$. The transition between this high-frequency limit and the steady-state limit occurs when $\omega \approx \omega_c$.

\section{Piezo-response of the FD devices}
\label{piezo}

\begin{figure}[t]
\includegraphics[clip,width=8.5 cm] {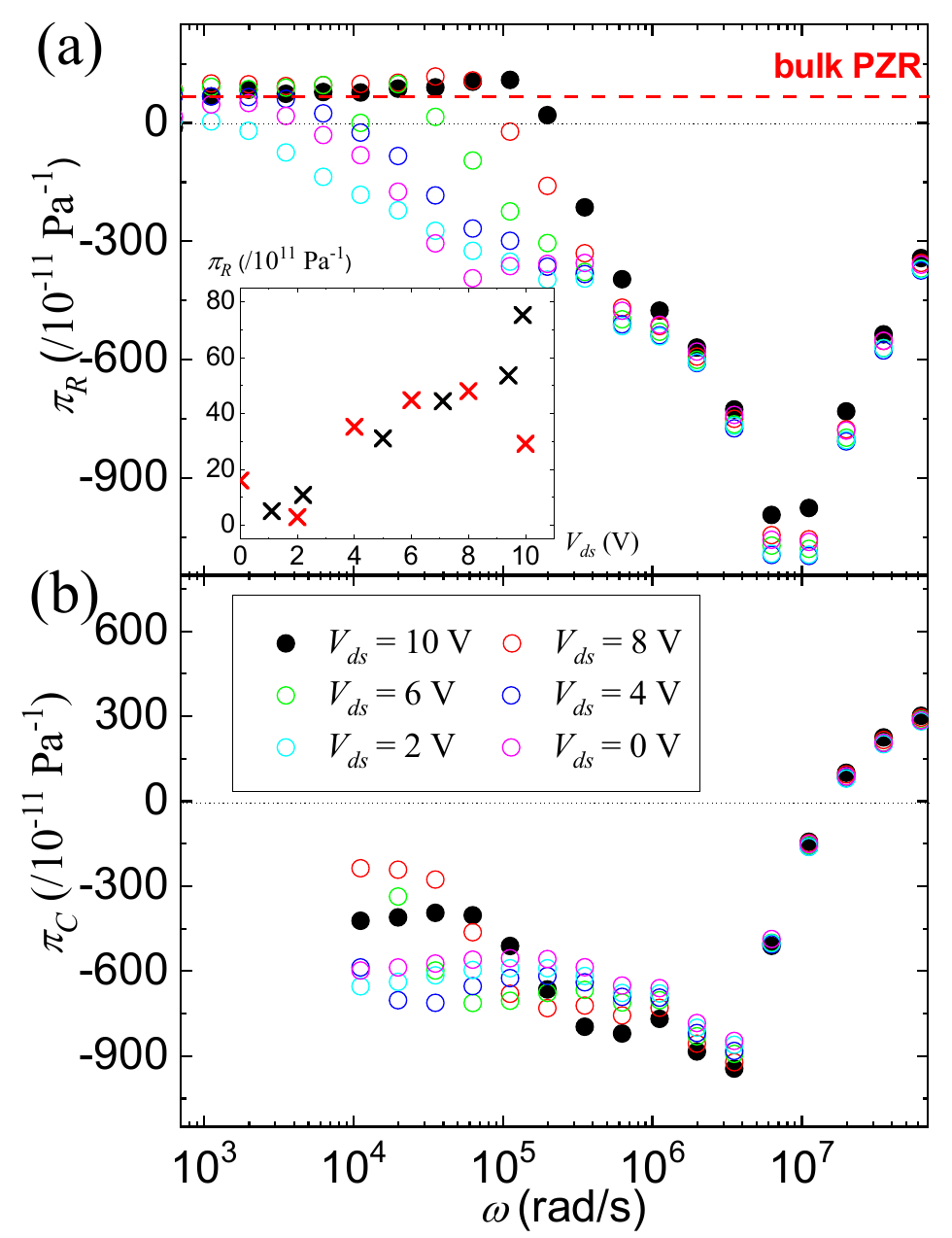}
\caption{(a) PZR and (b) PZC as a function of $\omega$ for several values of $V_{ds}$ obtained for an applied tensile stress of 25 MPa parallel to the $\langle 110 \rangle$ crystal direction. The sign and magnitude of these coefficients is to be compared with $\pi_{\textrm{bulk}}$ (red, dashed line in (a)). The inset in (a) compares the PZR $\pi$-coefficient measured using DC techniques (black crosses) with that obtained at low frequencies ($\omega < 1000$ rad/s with impedance spectroscopy (red crosses).}
\label{stress}
\end{figure}

A 25 MPa uni-axial tensile stress is now applied parallel to the $\langle 110 \rangle$ crystal direction. Typical results for $\pi_R$ and $\pi_C$ are shown in Fig. \ref{stress}. 

In order to link the impedance spectroscopy method with the DC techniques used previously, it is useful to begin by comparing the PZR measured in each case on the FD devices studied here. At the lowest frequencies where $\omega \ll \omega_c$ for all trap types shown in Table \ref{zerostressparams}, $\pi_R$ measured using impedance spectroscopy is only weakly dependent on $\omega$ as seen in Fig. \ref{stress}(a). A weak, quasi-linear dependence of $\pi_R$ on $V_{ds}$ is however apparent, and this is seen more clearly in the inset of Fig. \ref{stress}(a) where the red crosses correspond to the data points obtained at the lowest measurement frequency, $\omega = 1000$ rad/s. These data points are very similar to those obtained using the DC method outlined in Section \ref{Intro} and shown as black crosses in the inset of Fig. \ref{stress}(a). The good agreement between the two approaches not only increases confidence in the impedance spectroscopy piezo-response measurements, but reinforces the discussion in Section \ref{zero}, that at the lowest measurement frequencies used here the FD devices are indeed in the steady-state limit. 

In steady-state (i.e. $\omega \ll \omega_c$) therefore, $\pi_R$ is neither giant \cite{he2006,neuzil2010,yang2010,yang2011} or anomalous \cite{jang2014}. Aside from an intriguing (but weak) $V_{ds}$ dependence, it is rather similar in sign and magnitude to $\pi_{\textrm{bulk}}$ \cite{smith1954}, consistent with most previous steady-state PZR measurements on silicon nano-objects made using DC methods \cite{toriyama2003, reck2008, bui2009, mile2010, barwicz2010, milne2010, koumela2011, kumar2013, vietadao2015, mcclarty2016}. In the following it will be argued (using a combination of Kassing’s and Shockley-Read-Hall's (SRH) model \cite{shockley1952}) that when the Fermi level is pinned at a trap energy as indicated schematically in Fig. \ref{depletion}, and when recombination of injected electrons and holes is negligible, the steady-state PZR should be equal to $\pi_{\textrm{bulk}}$, as observed, in the SCLC regime in the presence of fast traps.

The principal effect of mechanical stress in solids is to shift electronic energy levels \cite{bardeen1950}; the aforementioned stress-induced effective mass change giving rise to bulk PZR of p-type silicon is due, for example, to shifts in the heavy- and light-hole valence band energies \cite{smith1954,milne2012}. Consider for simplicity only the case of electron capture and emission in the SRH model (similar expressions are valid for holes). The thermal emission rate is $\omega_e = C_n \exp\left[-\left(E_c-E_F\right)/k_BT\right]$ and, in steady state when recombination is neglected, $\omega_c n = \omega_e n_t$. Here $n$ is the density of electrons in the band, $n_t$ is the density of trapped electrons, and $C_n$ is the SRH capture constant \cite{shockley1952}. Using non-degenerate electron statistics the ratio appearing in Eq. (\ref{steadystate}) is then \begin{equation} \label{ratio} \frac{\omega_e}{\omega_c}=\exp\left[-2\left(E_F-E_T\right)/k_BT\right]. \end{equation} Here $E_c$ denotes the conduction band edge energy, $E_F$ the Fermi energy and $E_T$ the trap energy, all shown schematically in Fig. \ref{depletion}. Eq. (\ref{ratio}) shows explicity how stress-induced changes to $\omega_c$ and $\omega_e$ result from stress-induced changes to either or both of ($E_c-E_F$) and ($E_c-E_T$). However, according to Eq. (\ref{steadystate}), the observation of bulk-like steady-state PZR here suggests that while both $\omega_c$ and $\omega_e$ may change with stress, \textit{their ratio does not}. Using this constraint, it follows from Eq. (\ref{ratio}) that \begin{equation} \label{energies} \frac{dE_F}{dX}=\frac{dE_T}{dX}. \end{equation} This result is self-consistent with a pinning of $E_F$ at $E_T$ as shown schematically in Fig. \ref{depletion} for $W \gg t$. Interestingly, a stress dependence of $E_F$ pinned at $E_T$ was the basis for the “piezopinch” description of the giant, steady-state PZR in silicon nanowires \cite{rowe2008}, and a similar description was later evoked to describe unusual steady-state PZR of ultra-thin silicon layers \cite{yang2010,yang2011}. The impedance spectroscopy data, and the arguments presented here based on the SRH model, show that if recombination is negligible, in fact the opposite is true.

Having said this, the PZR does exhibit giant, anomalous behavior reaching $\pi_R \approx -1100\times 10^{-11}$ Pa$^{-1}$, but only at intermediate $\omega$ corresponding to non-steady-state conditions. The $\omega$-dependence of $\pi_R$ can be understood using a stress-dependent version of Kassing’s model in which $\omega_c$, $\omega_e$ and $T$ vary with stress. The relative variation of $T$ is assumed to be equal to that of the carrier mobility, yielding $dT/dX = -1.2 \times 10^{-18}$ s/Pa. This change is common to all six trap types. Stress-induced changes to $G_g$, $C_g$ and to the trap densities (i.e. the relative weights in Table \ref{zerostressparams}) are assumed to be negligible. The colored curves in Fig. \ref{Kassing}(c) correspond to the calculated PZR of each of the six trap types. The stress-dependence of $\omega_e$ and $\omega_c$ for each trap type are chosen in order that $\omega_e/\omega_c$ be stress-independent according to the arguments given above, and they are given in Table \ref{piezoparams}. The overall PZR (black curve in Fig. \ref{Kassing}(c)), calculated by summing the individual contributions, is well matched to data measured at $V_{ds} =$ 10 V. 

\begin{table}[t]
    \caption{The stress-dependence of the fitting parameters used with Kassing's model \cite{kassing1975} to calculate the $\omega$-dependence of the PZR and the PZC for $V_{\textrm{ds}}$ = 10 V in Fig. \ref{Kassing}(c) and Fig. \ref{Kassing}(d) respectively.}
    \centering
    \begin{tabular}{c|c|c|c}
      \textbf{Trap} & \textbf{Line color} & $d\omega_c/dX$ & $d\omega_e/dX$ \\
       \textbf{number} & \textbf{in Fig. \ref{Kassing}} & (rad/s/Pa) & (rad/s/Pa) \\
      \hline
      1 & orange & $-8.37\times 10^{-5}$ & $-6.52\times 10^{-4}$ \\
      2 & red & $-7.74\times 10^{-3}$ & $-7.74\times 10^{-6}$ \\
      3 & green & $-2.25\times 10^{-3}$ & $-8.5\times 10^{-7}$ \\
      4 & purple & $-4.75\times 10^{-2}$ & $-1.61\times 10^{-4}$ \\
      5 & brown & -0.12 & $-1.2\times 10^{-4}$ \\
      6 & cyan & -1 & $-1\times 10^{-2}$ \\
    \end{tabular}
  \label{piezoparams}
\end{table}

As for $G_0$, three PZR frequency regimes can be qualitatively described. (1) The steady-state limit, directly comparable to previous DC measurements as already discussed above. (2) The high frequency limit, $\omega \gg \omega_c$, where injected electrons and holes remain in the bands during the voltage period and the device behaves as if traps were absent. Any stress-induced changes to $\omega_c$ and $\omega_e$ are therefore irrelevant in this limit, and a relatively small PZR given by some combination of the bulk values for electrons and holes is expected. The measurement apparatus used here is not able to reach this frequency regime, although a significant drop in the PZR is observed at the highest frequencies in Fig. \ref{stress}. (3) In the intermediate range, $\omega \approx \omega_c$, only partial relaxation of the non-equilibrium injected charge from the bands to the traps occurs, and any stress-induced change to $\omega_c$ and $\omega_e$ results in large relative changes to the fraction of this charge which remains in the band during a voltage cycle. Thus the giant, anomalous PZR observed in this frequency regime is due to stress-induced changes to the charge capture dynamics at fast traps. In addition to the stress-induced change to the carrier mobilities, there is then an additional change in the non-equilibrium electron and hole densities present in the bands. This is a purely non-steady-state phenomena.

As shown in Fig. \ref{stress}(b), the PZC is another phenomenon revealed by impedance spectroscopy which is not accessible using DC methods. Unlike the steady-state PZR, the steady-state PZC is large, with $\pi_C \approx -600\times 10^{-11}$ Pa$^{-1}$ at $V_{ds} =$ 0 V. At higher frequencies it reaches approximately $\pi_C \approx -900\times 10^{-11}$ Pa$^{-1}$ before changing sign at the highest frequencies. In Kassing’s model there are no longer any free parameters available to determine the PZC; it must be consistent with the parameter values shown in Tables \ref{zerostressparams} and \ref{piezoparams} used to model the PZR. Summing of the individual PZC curves shown in Fig. \ref{Kassing}(d) for $V_{\textrm{ds}} = 10$ V can however be done with an arbitrary sign associated with each curve in order to account for the electron- or hole-like nature of the traps as follows. 

The capacitance is given by a ratio $\Delta Q/\Delta V$ where $\Delta Q$ is the incremental change in the charge stored in the FD device occuring due to an  incremental change in the potential difference from source to drain, $\Delta V$. Since the applied stress modifies the density of trapped charge according to the trapping dynamics arguments given above, this results in a change in the capacitance. The sign of the trapped charge (i.e. electrons or holes) will therefore determine the sign of the capacitance change with stress. If the trap is an electron trap, a stress induced reduction in the capture rate (see Table \ref{piezoparams}) will reduce the average density of trapped electrons resulting in a net positive change in the total trapped charge (and therefore an increase in the capacitance). According to the definition of $\pi_c$ in Eq. (\ref{piC}) this will yield a negative PZC coefficient. The opposite will be true of hole traps. Thus the individual PZC curves calculated using the parameters in Tables \ref{zerostressparams} and \ref{piezoparams} must be multiplied either by a factor of -1 for hole traps or +1 for electron traps. In Fig. \ref{Kassing}(d) the red (trap type number 2) and cyan (trap type number 6) curves have been multiplied by -1 and therefore nominally correspond to hole traps. The other four traps are therefore considered to be electron traps. If the fast traps involved here are the intrinsic Pb$_0$ interface defects (as the symmetry of the piezo-response discussed in Section \ref{piezosymmetry} suggests) this would be consistent with the amphoteric nature of such traps \cite{poindexter1984}. It is emphasized that since the conductivities of electrons and holes add, no such modification of the sign of the individual PZR components is necessary (or possible) in the calculation of the overall PZR. The resulting PZC sum yields an overall calculated $\pi_C$ at $V_{\textrm{ds}} =$ 10 V that is well matched to the experimental curve as shown in Fig. \ref{Kassing}(d).

Given the large number of model parameters and the limitations of Kassing's model discussed above, no claims can be made here as to the physical meaning of their values. On the other hand, the ability to simultaneously match all four experimental curves ($G_0$, $C_0$, $\pi_R$ and $\pi_C$) in Fig. \ref{Kassing} with a single set of parameters reinforces the interpretation based on stress-induced changes to the fast trap dynamics.

\begin{figure}[t]
\includegraphics[clip,width=8.5 cm] {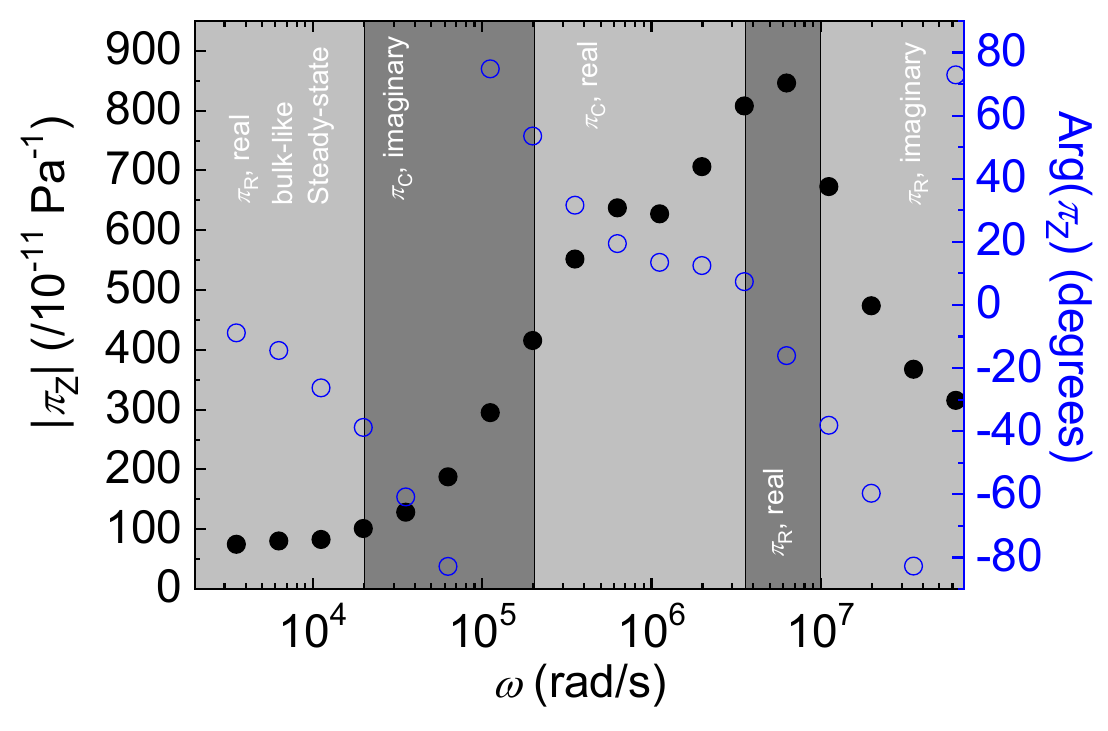}
\caption{The amplitude (black) and phase (blue) of $\pi_Z$ at $V_{ds}$ = 10 V calculated using Eq. \ref{piZ} from the data presented earlier. $\pi_Z$ can be real or imaginary, and may be dominated by either the PZR or the PZC depending on $\omega$ (see labels in gray boxes). In steady-state the overall piezo-response is close to the usual, bulk PZR given by $\pi_{\textrm{bulk}}$.}
\label{piezoimpedance}
\end{figure}

Using Eq. (\ref{piZ}) with the $G_0$, $C_0$, $\pi_R$ and $\pi_C$ data, $\pi_Z$ can be calculated as shown in Fig. \ref{piezoimpedance} for $V_{ds} =$ 10 V. At low and high frequency $\pi_Z$  is dominated by the PZR. It is imaginary at the highest frequencies, real at the peak piezo-response, and real at low frequencies (see gray boxes and labels). This result again emphasizes that in steady-state, the overall piezo-response is just the bulk PZR. At intermediate frequencies however, $\pi_Z$ is dominated by the PZC and switches from an imaginary to a real response with increasing frequency.

The interpretation of the giant, anomalous piezo-response as being due to stress-induced modifications of the fast trapping dynamics in the SCLC regime is further reinforced by two final observations; the strong, systematic voltage dependence of the PZR below $\approx 6\times 10^5$ rad/s shown in Fig. \ref{stress}(a), and the symmetry of the piezo-response in stress.

\section{Voltage dependence of the piezo-response}
\label{volts}

\begin{figure}[t]
\includegraphics[clip,width=8.5 cm] {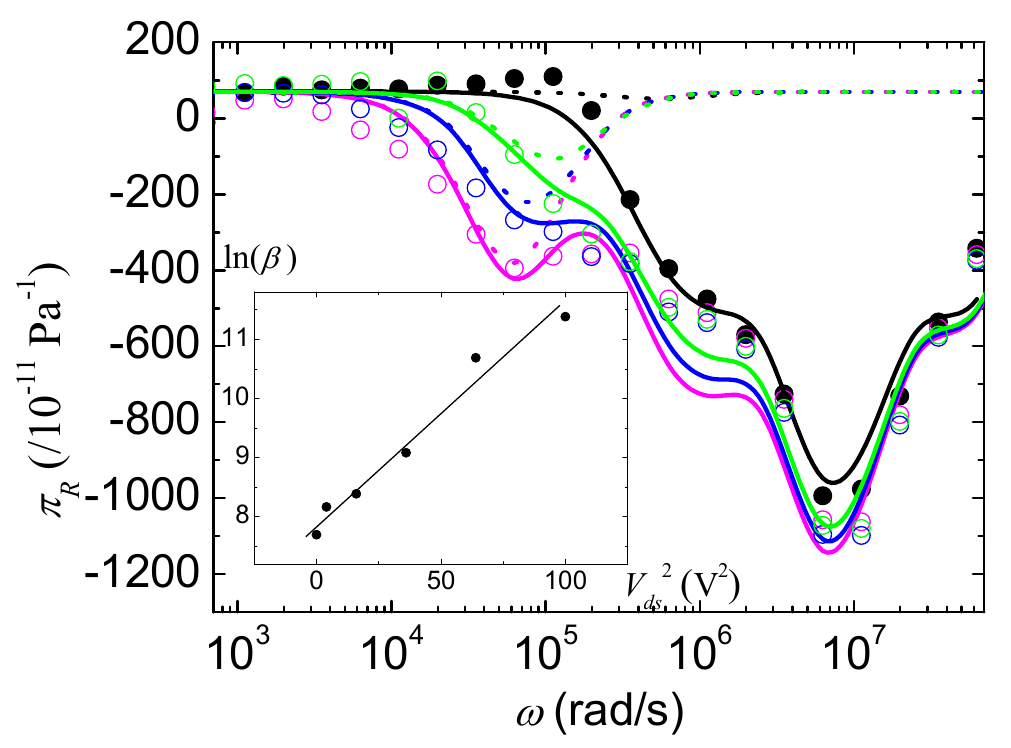}
\caption{$V_{ds}$ dependence of the PZR with the data (circles) shown according to the color scheme of Fig. \ref{stress}. An applied voltage increases all trap emission rates by a voltage dependent pre-factor $\beta$ whose values are chosen so as to best match the low frequency PZR. The resulting $V_{ds}^2$-dependence of $\ln\beta$ (see insert) suggests that the voltage dependence of the PZR results from changes to the trap emission rates arising phonon-assisted tunneling \cite{ganichev2000}.}
\label{Poole}
\end{figure}

It is known that $\omega_e$ can be affected by an applied electric field \cite{ganichev2000}, and it is tempting (but difficult) to attribute the $V_{ds}$-dependence of the PICTS signal in Fig. \ref{picts} of Appendix \ref{pictsappendix} to this. To account for the $V_{ds}$-dependence of the piezo-response, the emission rates for all six traps used here are allowed to vary with $V_{ds}$ by the same voltage dependent pre-factor, $\beta$\cite{ganichev2000}. Assuming a stress-independent $\beta$ then $d\omega_e/dX \rightarrow \beta d\omega_e/dX$. Since the low-frequency conductance is directly proportional to $\omega_e$ as in Eq. (\ref{steadystate}), the voltage dependence of the PZR is the relevant quantity to study. $\beta$ is chosen for each value of $V_{ds}$ (see Table \ref{beta}) in order to best match the $V_{ds}$-dependence of the lowest frequency PZR peak (dashed lines for four $V_{ds}$ values in Fig. \ref{Poole}). The resulting overall PZR is shown as solid lines whose color corresponds to that of the data (circles) in Fig. \ref{stress}. This procedure predicts not only the $V_{ds}$-dependence of the lowest frequency PZR peak, but also the relative insensitivity of the higher frequency PZR to changes in $V_{ds}$, despite the fact that the emission rates of the higher capture rate traps have also been multiplied by $\beta$. 

As can be seen from the values of $\omega_c$ and $\omega_e$ in Table \ref{zerostressparams}, the relative sensitivity of the low frequency measurements to $V_{ds}$ arises because the trap with the lowest capture rate (orange lines in Fig. \ref{Kassing}) transitions from being a fast trap at $V_{ds}$ = 0 V to a slow trap at higher voltages. To make this explicit, consider trap type number 1 whose emission rate at $V_{ds}=0$ V is 87750/40 = 2193 rad/s according to the values of $\omega_e$ in Table \ref{zerostressparams} and $\beta$ in Table \ref{beta}. This is less than the capture rate (16875 rad/s) which is voltage independent, and so type 1 traps are fast trap at $V_{ds}=0$ V. It transits from being a fast to a slow trap for an applied bias between 6 V and 8 V. Since only fast traps contain non-negligible charge concentrations and therefore contribute to the anomalous piezo-response, this provides an explanation for the voltage dependence of the type 1 trap contribution to the PZR. By combining the values from Tables \ref{zerostressparams} and \ref{beta}, the reader will see that all the other traps are fast over the experimentally tested range of $V_{ds}$. This analysis can be pushed further by considering the inset of Fig. \ref{Poole} in which $\ln\beta$ is plotted against $V_{ds}^2$. The observed linear variation rules out the Poole-Frenkel effect and favours a phonon-assisted tunneling emission process \cite{ganichev2000}.

\begin{table}[t]
    \caption{Voltage dependent pre-factors, $\beta$, used to account for the increase in all trap emission rates with increasing applied voltage.}
    \centering
    \begin{tabular}{c|c|c|c|c|c|c}
      \textbf{$V_{\textrm{ds}}$} (V) & 10 & 8 & 6 & 4 & 2 & 0 \\
      \hline
      $\beta$ & 40 & 20 & 4 & 2 & 1.6 & 1 \\
    \end{tabular}
  \label{beta}
\end{table}

\section{Symmetry of the piezo-response}
\label{piezosymmetry}

\begin{figure}[t]
\includegraphics[clip,width=8.5 cm] {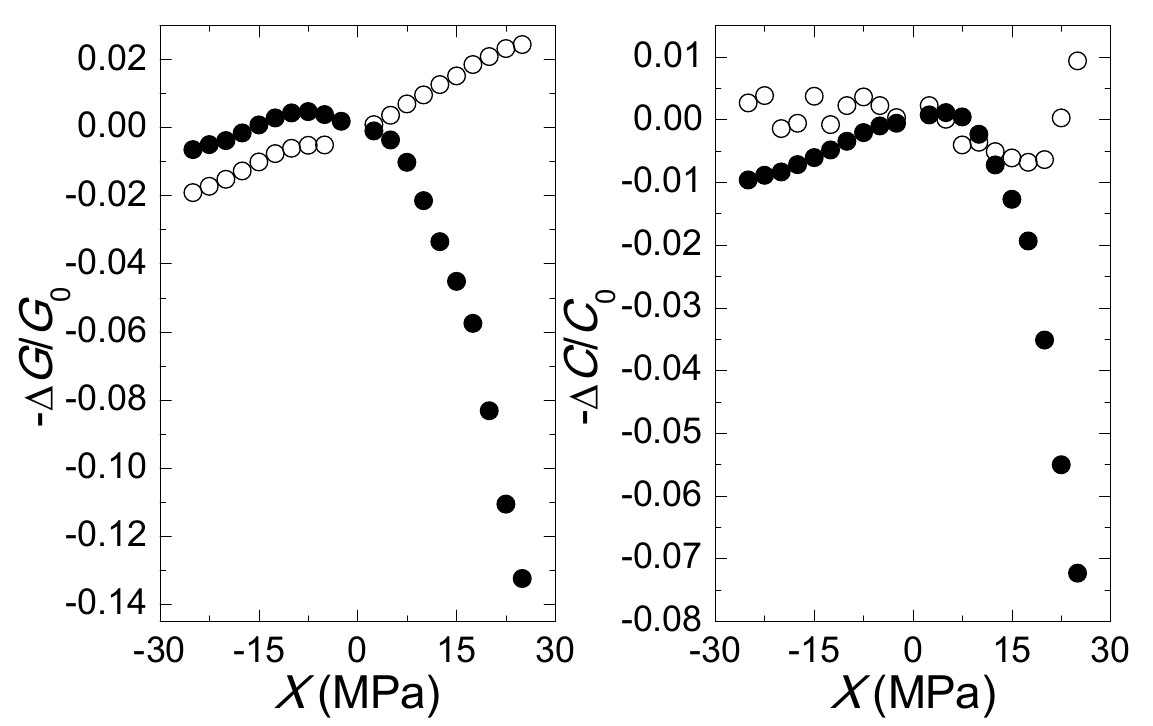}
\caption{Stress symmetry of the PZR and the PZC at $\omega = 1.5\times 10^4$ rad/s (open circles) and at $\omega = 1.5\times 10^7$ rad/s (filled circles) for $V_{ds}$ = 10 V. An odd symmetry is consistent with the usual bulk PZR, while an even symmetry can be attributed to electro-mechanical activity of intrinsic Pb$_0$ centers at the silicon/oxide interface.}
\label{symmetry}
\end{figure}

Figure \ref{symmetry} shows the relative stress-induced changes in $G$ and $C$ at two measurement frequencies, $\omega = 1.5\times 10^4$ rad/s (open circles) and $\omega = 1.5\times 10^7$ rad/s (filled circles) for $V_{ds}$ = 10 V. Negative values of $X$ correspond to compression. At low frequency $-\Delta G/G_0$ is approximately linear (i.e. odd) in stress as would be expected for the steady-state, bulk PZR \cite{smith1954}. The stress symmetry of the PZC at low frequency is not clear, partly because capacitance measurements are difficult at such frequencies. Unlike the PZR however, it is not clearly odd. The high frequency PZR and PZC are both even (but asymmetric) in stress. This symmetry has been reported elsewhere, including in the gate leakage currents \cite{choi2008} and in flash EEPROMs \cite{toda2005}, and has even been reported in steady-state PZR measurements made using DC techniques on depleted silicon membranes \cite{yang2010,yang2011}. More recently, spectroscopic studies of the stress-dependence of the surface band bending at a silicon/oxide interface was also found to be even in stress \cite{li2018b}. In some cases \cite{toda2005,choi2008,li2018b} the symmetry of the Pb$_0$ center at the silicon/oxide interface is invoked to explain the even response, so Fig. \ref{symmetry} provides tentative evidence that these defects are the electro-mechanically active fast traps responsible the non-steady-state piezo-response. This conclusion is also consistent with the broad PICTS signal shown and discussed in Appendix \ref{pictsappendix} (see Fig. \ref{picts}).

\section{Conclusions}

Using impedance spectroscopy techniques the steady-state piezo-response of the SCLC in natively-oxidized silicon is found to be just the usual bulk PZR. This reinforces the majority of reports of such behavior in depleted, silicon nano-objects \cite{toriyama2003, reck2008, bui2009, mile2010, milne2010, barwicz2010, koumela2011, kumar2013, vietadao2015, mcclarty2016}. Using a combination of the SRH model and Kassing's model for small-signal SCLC, it is argued that this is a consequence of surface Fermi level pinning by silicon/oxide interface defects, in direct opposition to previous models of giant PZR \cite{rowe2008}. The observations suggest possible explanations for discrepancies in the nano-silicon PZR literature, in particular reports of giant or anomalous effects at small applied stresses in the steady-state \cite{he2006, neuzil2010, kang2012, yang2010,yang2011, jang2014}. One possibility is that in the nano-objects in which giant or anomalous PZR is reported, $E_F$ was not pinned at $E_T$ so that Eqs. (\ref{ratio}) and (\ref{energies}) are no longer valid. At first sight this seems surprising since pinning of the Fermi level at the silicon/oxide interface is well known, especially for native oxides \cite{hollinger1983}. However, in some of these reports chemical surface treatments were used, which may have significantly modified the surface electronic structure \cite{he2006,yang2010,yang2011,jang2014}. A second possibility is that recombination and its stress dependence were more important in these works than in the FD-SOI wafers studied here. If so, then Kassing's model used here is no longer applicable, and the full equations-of-motion given in Appendix \ref{KassingEquations} must be used to describe the small-signal SCLC and its piezo-response. Another possibility is that in reports of giant or anomalous PZR, the measurements were not strictly made in the steady-state. If the I-V characteristic was obtained by rapidly sweeping the applied voltage with a source-measure unit, traps may no longer be able to reach steady-state \cite{bayliss2005}, and the apparent steady-state PZR may in fact be a mixture of the high-frequency PZR and the PZC reported here.

Impedance spectroscopy methods are also shown to give access to the non-steady-state piezo-response where giant, anomalous PZR is observed. In addition, the quadrature response is found to correspond to a giant PZC. Using a simplified, stress-dependent model of the SCLC in the presence of fast traps, and by observing the voltage-dependence of the PZR, these phenomena are shown to be the result of electro-mechanically active fast traps whose capture and emission rates are stress-dependent. In this case at measurement frequencies $\omega \approx \omega_c$, the PZR consists of not only the stress-induced change in the charge carrier mobilities \cite{smith1954} i.e. the bulk effect, but also a significantly larger stress-induced change in the concentration of non-equilibrium, quasi-free electrons and holes injected from the ohmic contacts. The PZC is in some respects complementary to the PZR in that it depends on stress-induced changes to the \textit{trapped} electron and hole concentrations. The symmetry of the the giant, anomalous PZR and the PZC, suggest that intrinsic Pb$_0$ silicon/oxide interface defects are the likely candidates for the electro-mechanically active fast traps.

These observations suggest a number of interesting experiments, including stress-dependent defect spectrocopy to evaluate the stress-induced changes in $E_T$, studies of the piezo-response of FD-device into which specific, well-chosen defects have been engineered, studies of the piezo-response of unipolar devices where recombination is certainly absent, studies of the non-steady-state piezoresponse of piezo-resistive, nano-mechanical oscillators whose resonant frequency is comparable with $\omega_c$ \cite{mile2010}, or studies of the piezo-response of transistors in the sub-threshold region where SCLCs occur and which are of interest for ultra-low power consumption applications \cite{wang2006, kang2012}. Moreover, in some proposed future, quantum devices, fast traps are desirable for device operation \cite{mccamey2006} while in other cases SCLCs are difficult to avoid, for example in organic materials \cite{alison1994,leroy2003,mark1999}. From these examples it is clear that the ability to significantly modify trap-mediated SCLCs with mechanical stress must at the very least be accounted for, and may in fact provide a route to new functionalities.

\acknowledgements{This work was partially financed by the French Agence Nationale de la Recherche, contract ANR-17-CE24-0005.}

\appendix
\section{Photo-induced current transient spectroscopy}
\label{pictsappendix}

\begin{figure*}[t]
\includegraphics[clip,width=16 cm] {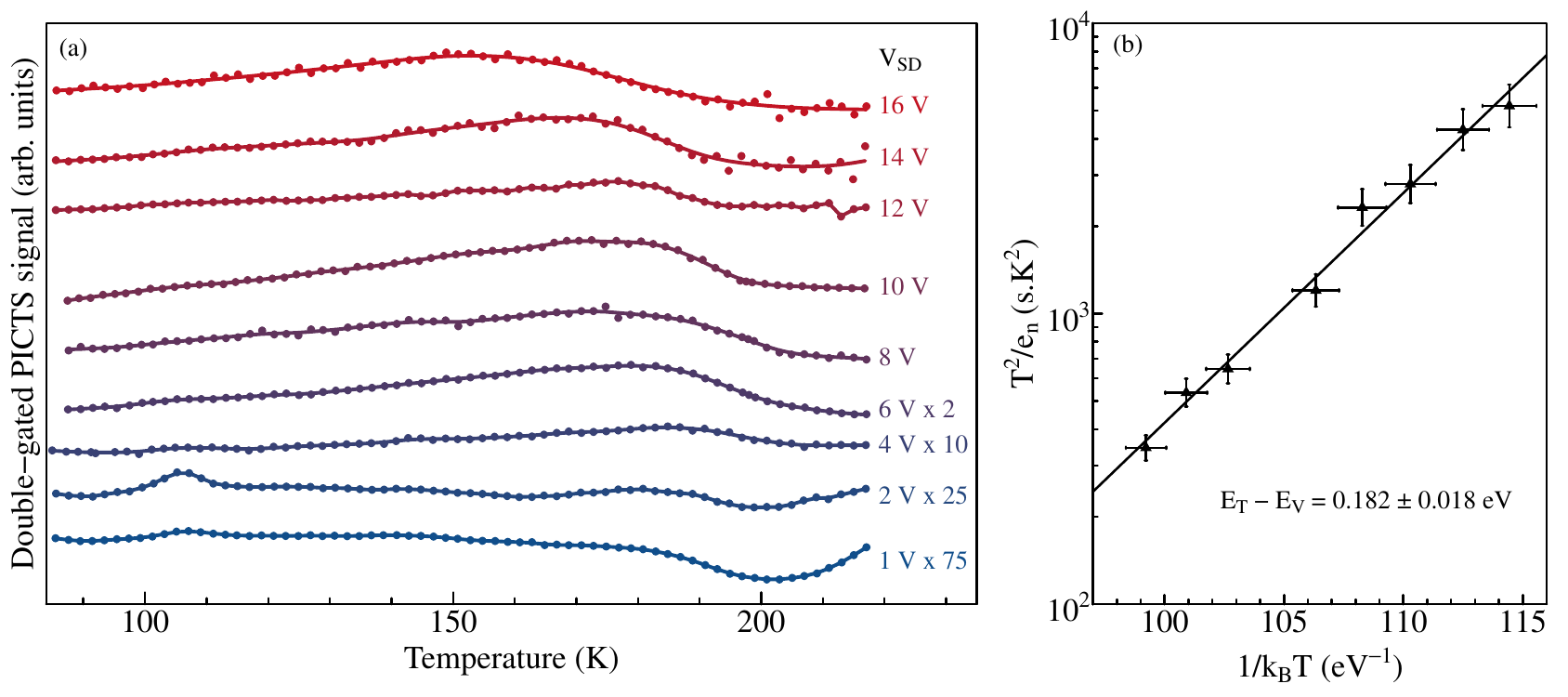}
\caption{(a) The PICTS signal obtained as a function of $V_{ds}$ displaced vertically for clarity. A smoothing spline fit is fitted to the raw PICTS data as a guide the reader’s eye. The asymmetric, extended hump visible for $V_{ds} \geq 4$ V is usually associated with a continuous trap distribution which may be associated silicon/oxide interface states \cite{papaioannou1989}. (b) An Arrhenius plot constructed from a set of PICTS signals obtained using different $t_1$ and $t_2$ times reveals that the sharp structure visible at $V_{ds} = 1$ V in (a) corresponds to a trap lying approximately 0.182 eV above the valence band edge.}
\label{picts}
\end{figure*}

Since the impedance spectroscopy data suggests that both the zero-stress and piezo-response properties of the FD devices are determined by the electro-mechanical properties of fast electronic traps, initial attempts at defect spectroscopy measurements were made on the FD devices using photo-induced current transient spectroscopy (PICTS). PICTS is a varient of deep level transient spectroscopy (DLTS) \cite{lang1974} which is used to investigate deep level traps in high-resistivity materials and devices \cite{papaioannou1989,balland1986,balland1986b}. By utilizing a periodic above bandgap optical excitation to generate electron-hole pairs, traps are readily filled by photocarriers. Immediately after the end of the optical excitation, a sharp current drop due to recombination is observed, followed by a slower delay due to trapped charge carrier emission. By monitoring the detrapping of trapped charge carriers as a function of temperature, important parameters about the traps present can be determined. 

Here, PICTS signals were obtained using a pulsed 940 nm commercial, high-speed light emitting diode with a rise and fall time of 20 ns. A signal generator (Agilent 33210A) was used to apply a 100 ms pulse to the emitting diode at a rate of 1 Hz, which delivered an optical pulse of approximately 55 mW to the sample. In the only PICTS data presented here, $V_{ds}$ is fixed throughout the temperature scan. The resulting photo-current transient is measured using a custom built DLTS setup consisting of a SR 570 current amplifier with the temperature ramped from 86 K to 300 K in 2 K increments. At each temperature step 30 transients were averaged in order to improve the signal-to-noise ratio. The averaged current transients are then processed using a DLTS double boxcar analysis with $t_1$ and $t_2$ times chosen according to $t_2/t_1 = 2$ and $t_1 = 175$ ms to obtain the PICTS signals shown in Fig. \ref{picts}(a).

Figure \ref{picts}(a) shows the PICTS signal as a function of $V_{ds}$ displaced vertically with increasing voltage for clarity. For $V_{ds}=1$ V a clear peak centered about 106 K can be seen. The Arrhenius plot in Fig. \ref{picts}(b) is constructed from a set of PICTS signals obtained using different $t_1$ and $t_2$ times, and reveals the trapping level lying approximately 0.182 eV above the valence band edge. The origin of this discrete trap level is unclear. As $V_{ds}$ is increased, the peak at approximately 106 K disappears. Instead, an asymmetric broad hump centered about 180 K is observed. This continuous non-zero distribution is usually associated with interface states described by a continuum of states within the band gap \cite{papaioannou1989}. It is also noted that changes to $V_{ds}$ significantly shift the broad PICTS signal in temperature, and that a change in the shape of the broad distribution is also apparent. It is tempting to ascribe this variation either to the Poole-Frenkel effect or to phonon-assisted tunneling emission from traps \cite{ganichev2000}, but no clear tendency (for example in the peak PICTS signal position, or in the characteristic PICTS relaxation times) that would permit an identification of either phenomenon is evident from the data.

\begin{figure}[t]
\includegraphics[clip,width=8.5 cm] {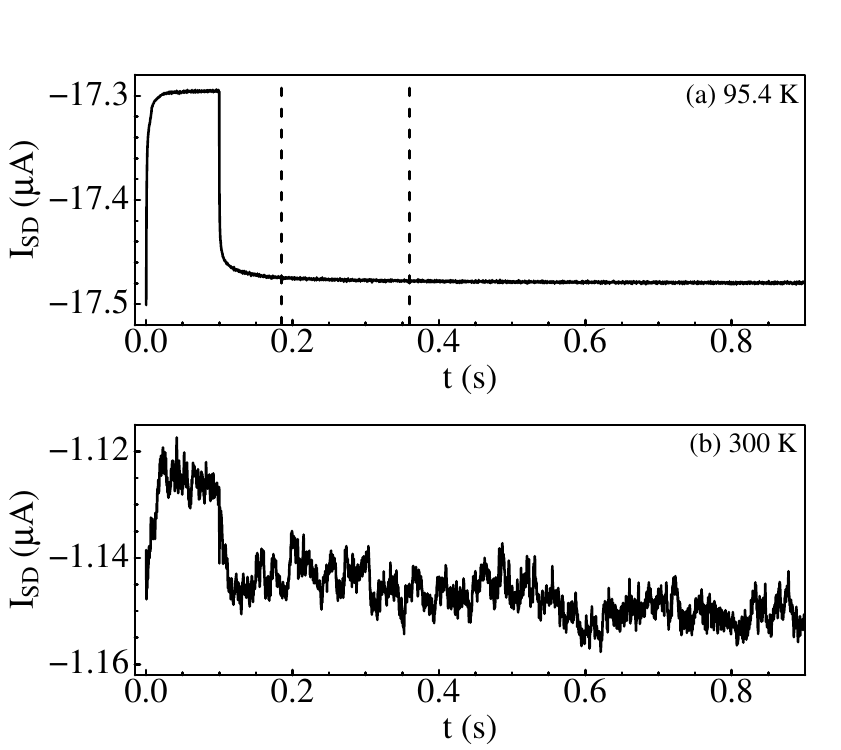}
\caption{Raw photo-induced current transients measured at (a) 101.3 K and (b) 300.0 K with $V_{ds} = 6$ V. The clean transient in (a) permits a double boxcar analysis (dotted lines) that is used to construct the PICTS signal in Fig. \ref{picts}(a). At elevated temperatures in (b), the amplitude of the random telegraph noise is comparable to the photo-induced current transient. Boxcar analysis at high temperatures is therefore avoided.}
\label{telegraph}
\end{figure}

At elevated temperatures, current transient fluctuations as shown in Fig. \ref{telegraph}, likely due to random telegraph noise, start to dominate the current transient. As the random telegraph noise amplitude is comparable to the current transient amplitude, double boxcar analysis used for PICTS is not appropriate. Therefore, useful PICTS signals are restricted to the temperature range from 90 K to 220 K, although measurements up to room temperature were performed. Note also that in the room temperature piezo-response measurements reported in the main manuscript, the random telegraph noise is averaged out so that only the mean value of the conductance and capacitance are considered.

\section{Limitations of Kassing's model}
\label{KassingEquations}

As pointed out in Section \ref{zero}, Kassing's model \cite{kassing1975}, which gives a small-signal analysis of a SCLC in the presence of traps, treats only a single charge carrier type (electrons or holes but not both) and therefore has (in principal) a limited applicability in bipolar SCLC devices like those studied here. The full equations of motion which describe bipolar SCLC transport in the presence of traps are given by the following time-dependent, coupled differential equations in 1-dimension:

\begin{widetext}
\begin{align} 
E(x,t) &= -\frac{\partial V(x,t)}{\partial x}, \label{field} \\ 
J(x,t) &= (n(x,t)\mu_nq+p(x,t)\mu_pq)E(x,t) + \epsilon_r\epsilon_0 \frac{\partial E(x,t)}{\partial t}, \label{Jcurrent} \\ 
\frac{\partial E(x,t)}{\partial x} &= -\frac{q}{\epsilon_r\epsilon_0}\left[p_t(x,t)+p(x,t)-n_t(x,t)-n(x,t)-p_{t0}-p_0+n_{t0}+n_0 \right], \label{poisson} \\ 
\frac{\partial n(x,t)}{\partial t} &= -C_n \left \{ n(x,t)\left[ \frac{N_t-n(x,t)}{N_t} \right] - \frac{n_1}{N_t}n_t(x,t) \right \} + \frac{1}{q}\frac{\partial J_n(x,t)}{\partial x}, \label{electrons} \\ 
\frac{\partial p(x,t)}{\partial t} &= -C_p \left \{ p(x,t)\frac{n_t(x,t)}{N_t}  - \left[ \frac{N_t-n_t(x,t)}{N_t} \right] \right \} - \frac{1}{q}\frac{\partial J_p(x,t)}{\partial x}, \label{holes} \\
\frac{\partial n_t(x,t)}{\partial t} &= C_n \left \{ n(x,t)\left[ \frac{N_t-n(x,t)}{N_t} \right] - \frac{n_1}{N_t}n_t(x,t) \right \} -S_0n_t(x,t)p_t(x,t), \label{trappedelectrons} \\ 
\frac{\partial p_t(x,t)}{\partial t} &= C_p \left \{ p(x,t)\frac{n_t(x,t)}{N_t}  - \left[ \frac{N_t-n_t(x,t)}{N_t} \right] \right \} -S_0n_t(x,t)p_t(x,t). \label{trappedholes} \end{align}
\end{widetext}

In this set of equations several terms are recognizable. In Eq. (\ref{Jcurrent}) for the overall current density, $J_n(x,t) = n(x,t)\mu_nqE(x,t)$ and $J_p(x,t) = p(x,t)\mu_pqE(x,t)$ are the electron and hole current densities appearing respectively in the rate equations, Eq. (\ref{electrons}) and Eq. (\ref{holes}). Here $q$ is the elemental electronic charge, $\mu_n$ and $\mu_p$ the electron and hole mobilities respectively, $n(x,t)$ and $p(x,t)$ the quasi-free electron and hole densities respectively, and $E(x,t)$ the electric field. The second term on the right-hand-side of Eq. (\ref{Jcurrent}) is the displacement current with $\epsilon_r\epsilon_0$ the permittivity of the medium. In the rate equations for free electrons and holes, Eq. (\ref{electrons}) and Eq. (\ref{holes}) respectively, the terms in the curly brackets are the net capture rates of electrons and holes at traps of density $N_t$, as originally defined by Shockley and Read \cite{shockley1952}. $C_n$ and $C_p$ are the electron and hole capture coefficients, and $n_t(x,t)$ and $p_t(x,t)$ are the trapped electron and hole densities respectively. $n_1$ and $p_1$ are the equilibrium electron and hole densities that would be obtained with $E_F = E_T$. In the rate equations for the trapped electrons and holes, Eq. (\ref{trappedelectrons}) and Eq. (\ref{trappedholes}) respectively, the terms in curly brackets are once again the electron and hole capture rates, and the terms proportional to the product $n_t(x,t)p_t(x,t)$ represent recombination. $S_0$ is the so-called bi-molecular recombination coefficient. The remaining variables appearing in Poisson's equation, Eq. (\ref{poisson}), are the equilibrium concentrations of trapped holes ($p_{t0}$), free holes ($p_0$), trapped electrons ($n_{t0}$) and free electrons ($n_0$).

It appears that these equations have not been fully treated in the literature, at least in the small-signal limit where it is only possible to find unipolar treatments either without \cite{shao1961,vanderziel1966,dascalu1966,wright1966, kassing1975} or with a single trapping level \cite{many1962,dascalu1966,kassing1975}. The one small-signal, bipolar treatment that includes a single trapping level makes a number of assumptions regarding the relative trapping, re-emission and recombination rates which renders it a quasi-steady-state model \cite{nicolet1975}. It cannot describe the impedance spectroscopy data presented here in Figs. \ref{nostress}, \ref{Kassing} and \ref{stress}. More recently, similar sets of equations have been used to model charge leakage in organic materials but in these cases simplifying assumptions are still made; in some cases the trapping and re-emission expressions are not explicitly consistent with the non-steady-state capture rates given in the SRH model \cite{alison1994,leroy2003}, whereas in others recombination is assumed to be infinitely fast \cite{pitarch2006}. In short, the solution of Eqns. (\ref{field}--\ref{trappedholes}) remains an outstanding challenge, but is in principal necessary to properly evaluate the relative importance of the stress-dependence of the capture, re-emission and recombination rates.

As already presented in Sections \ref{zero} and \ref{piezo}, in lieu of a full resolution of Eqns. (\ref{field}--\ref{trappedholes}), Kassing's model \cite{kassing1975} is used. In this approach a reduced (unipolar) set of equations is then written as follows:

\begin{widetext}
\begin{align} 
E(x,t) &= -\frac{\partial V(x,t)}{\partial x}, \label{field2} \\ 
J(x,t) &= n(x,t)\mu_nqE(x,t) + \epsilon_r\epsilon_0 \frac{\partial E(x,t)}{\partial t}, \label{Jcurrent2} \\ 
\frac{\partial E(x,t)}{\partial x} &= -\frac{q}{\epsilon_r\epsilon_0}\left[-n_t(x,t)-n(x,t)+n_{t0}+n_0 \right], \label{poisson2} \\ 
\frac{\partial n(x,t)}{\partial t} &= -C_n \left \{ n(x,t)\left[ \frac{N_t-n(x,t)}{N_t} \right] - \frac{n_1}{N_t}n_t(x,t) \right \} + \frac{1}{q}\frac{\partial J(x,t)}{\partial x}, \label{electrons2} \\ 
\frac{\partial n_t(x,t)}{\partial t} &= C_n \left \{ n(x,t)\left[ \frac{N_t-n(x,t)}{N_t} \right] - \frac{n_1}{N_t}n_t(x,t) \right \}. \label{trappedelectrons2}  \end{align}
\end{widetext}
 
These equations have been written for electrons, but may equally well have been written for holes. Kassing explicitly defines the parameters $\omega_c$ and $\omega_e$ used in Sections \ref{zero} and \ref{piezo} in terms of the SRH capture constant and the trap density: \begin{equation} \label{kassingomegac} \omega_c = C_n\frac{N_T-n_{T0}}{N_T}, \end{equation} and \begin{equation} \label{kassingomegae} \omega_e = C_n\frac{n_0+n_1}{N_T}, \end{equation} where $n_{T0}$ is the equilibrium trapped electron density, and then finds a complex, analytical solution for the admittance, $Y = J/E$, which is used throughout this work. It is explicitly seen that the recombination term is absent so that any dependence of $S_0$ in Eq. (\ref{trappedelectrons}) and Eq. (\ref{trappedholes}) on mechanical stress is lost in Kassing's model. Moreover, Eq. (\ref{kassingomegac}) and Eq. (\ref{kassingomegae}), show that the attribution of a stress-dependence to $\omega_c$ and to $\omega_e$ as in Table \ref{piezoparams}, is in fact better ascribed to a stress-dependence of $C_n$ from the SRH model, and to $n_1$. The fact however, that Kassing's model is able to reproduce relatively well the observed dependences of $G_0$, $C_0$, $\pi_R$ and $\pi_C$ on $\omega$ tentatively suggests that the stress-dependence of the bi-molecular recombination constant, $S_0$, is of secondary importance. It also tentatively suggests that electrostatic coupling between electrons and holes via Poisson's equation, Eq. (\ref{poisson}), is also to a certain extent negligible. A better evaluation of these conclusions can only be made by numerically solving Eqns. (\ref{field}--\ref{trappedholes}) which will be the object of future work.

\bibliographystyle{apsrev}

\bibliography{F:/Publications/References}
\end{document}